\documentclass[a4paper,10pt]{article}
\usepackage{graphicx}

\begin{document}

  \title{Calibration and Optimization of a Very Large Volume Neutrino Telescope using Extensive Air Showers\footnote{Published in NIMA, doi:
10.1016/j.nima.2010.06.247, representing the KM3NeT Consortium}}
 \author{A. Leisos\footnote{Physics Laboratory, School of Science \& Technology, Hellenic Open University, Tsamadou 13-15 \& Ag. Andreou, Patras 26222, Greece},G. Bourlis, N.A.B. Gizani, E.P. Christopoulou\footnote{Astronomical Laboratory, Department of Physics, University of Patras, University campus, Rio 26504, Greece}, A.G. Tsirigotis\\ \& S.E. Tzamarias}

\date{}
\maketitle

\paragraph{Abstract}

We report on a simulation study of the calibration potential offered by floating Extensive Air Shower (EAS) detector stations (HELYCON), operating in coincidence with the KM3NeT Mediterranean deep-sea neutrino telescope. We describe strategies in order to investigate for possible systematic errors in reconstructing the direction of energetic muons as well as to determine the absolute position of the underwater detector.\\


\begin{figure}
\begin{center}
\includegraphics*[width=9cm]{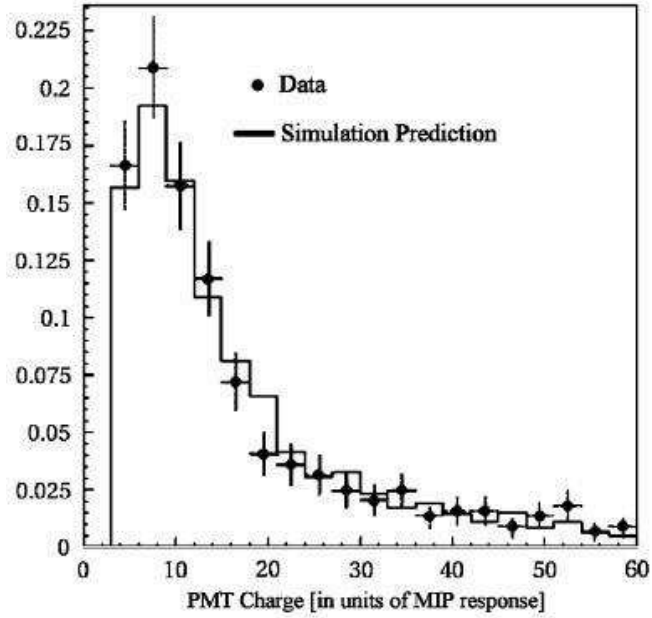}
\end{center}
\caption{ The collected total charge as a function of the distance, at sea level, of the shower axis from the center of the platform. The points correspond to the average total charge collected when at least one HELYCON detector is active (triangles) and when at least five detectors had responded in coincidence (circles). The error bars correspond to the RMS of the charge distribution per radial distance bin (3 m). The insert plot represents the radial distribution of selected EAS when at least five detectors were active with a collective response corresponding to more than 25 mip.}
\label{figcharge}
\end{figure}
\begin{figure}
\begin{center}
\includegraphics*[width=8cm]{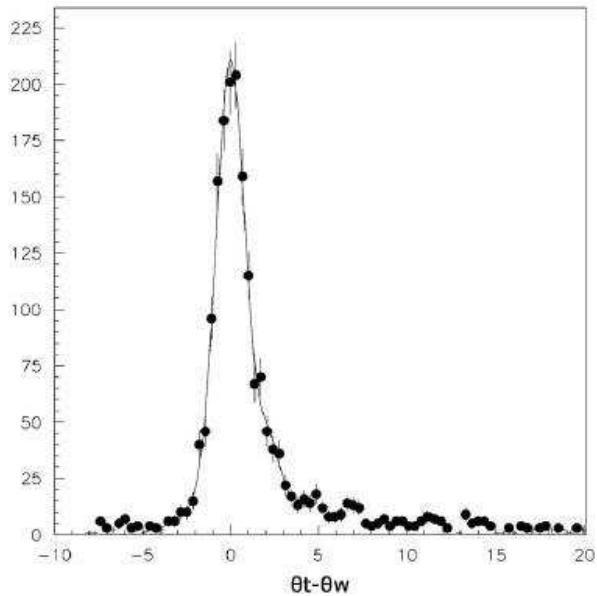}
\end{center}
\caption{The distribution of  the difference of the two zenith angle estimations, corresponding to 39 hours of data taking.}
\label{fig1}
\end{figure}

\section{Introduction}
Energetic atmospheric muons, produced in secondary processes during the Extensive Air Shower (EAS) development, reach an underwater neutrino telescope from above with a rate exceeding in several orders of magnitude any neutrino source\cite{Nestor,apothesis}. With previous studies we have shown \cite{apocalib,ernen,idm} that reconstructed EAS by floating HELYCON detectors can be used to investigate possible systematic errors. In this paper we report on the development of a new calibration strategy and we study its potential by means of Monte Carlo experimentation.
\section{Description of Calibration Strategies}
\label{EG}
The calibration methods we propose are based on the simultaneous detection of an EAS, on the sea-top of the telescope, and of at least one energetic muon by the neutrino telescope. We consider that the EAS will be detected by three independent HELYCON detector arrays \cite{idm}, positioned on floating platforms above the neutrino telescope. Each platform will carry 16 HELYCON charged particle detectors, each of 1$m^{2}$ effective area,  arranged on a two dimensional grid (5m cell size) covering a total area of about 360$ m^{2}$. The detectors on each platform will form an autonomous array equipped with a GPS antenna, for absolute synchronization and positioning, digitization and control electronics, as well as a data acquisition system controlled by a personal computer \cite{bourlistot,bouthesis}.

In our previous studies\cite{apocalib,ernen,cdr} the recorded signals by a HELYCON autonomous array were used to reconstruct the direction of the EAS axis and the coordinates of the point where the axis intercepts the horizontal plane at sea level.  The evaluation of a possible angular offset and the neutrino telescope position were achieved by comparing the shower axis parameters with the corresponding muon track parameters, reconstructed by the neutrino telescope\cite{apocalib}. 

\begin{figure}
\begin{center}
\includegraphics*[width=8cm]{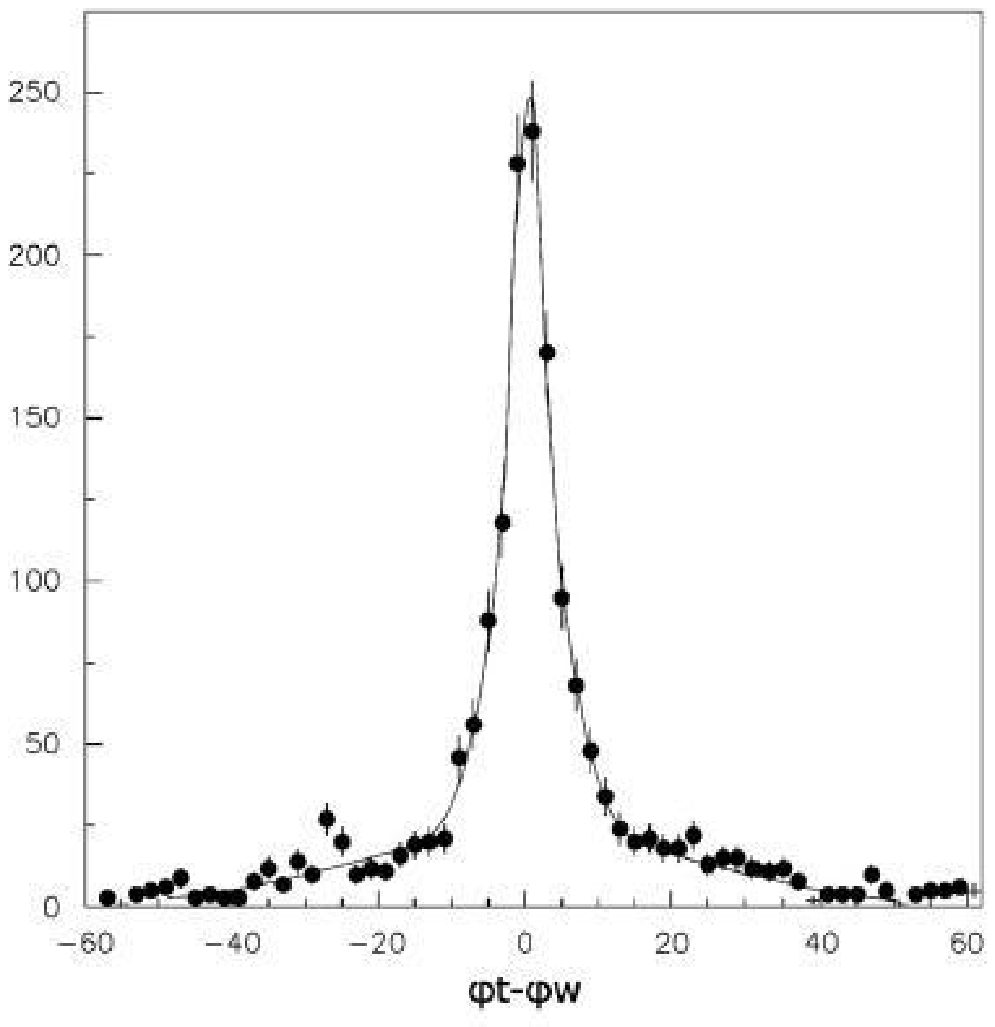}
\end{center}
\caption{The distribution of  the difference of the two azimuth angle estimations, corresponding to 39 hours of data taking.}
\label{fig3}
\end{figure}

Alternatively, each HELYCON autonomous array can be used to select EAS passing close to the center of the floating platform. When the telescope has detected at the same time (within a proper time widow) a down-coming muon track, a simple estimation of the muon track direction can be made \cite{gaisser}. That is the straight line (henceforth simple estimation) connecting the position of the center of the platform with the weighted mean (weighted by the observed charge) of the active\footnote{These are the optical modules selected by the filtering algorithm to define the muon track \cite{apohere}.}  optical modules positions. Simulation studies have shown that by requiring several synchronously active HELYCON detectors in the array, with a collective response corresponding to many ionizing particles, the selection favors EAS passing very close to the platform. 
 Figure \ref{figcharge} presents the total charge, collected by the active detectors of the array, as a function of the  radial distance of the shower axis,  at  sea level, from the center of the platform. When requiring five or more detectors to respond in coincidence and the total collected charge to exceed the equivalent of 25 mips, the average radial distance is 44 m, whilst only 0.05$\%$ of the selected EAS are farther than 150m from the center of the platform.  

Consequently, this simple estimation offers a good zenith angle resolution which is, according to our simulation studies, better than $0.7^{o}$ and also it is free of reconstruction bias. The simple estimation of the muon direction is compared, on an event by event basis, to the fully reconstructed  muon track by the neutrino telescope (henceforth reconstructed muon track). Figure \ref{fig1} presents the distribution of the difference between the simple estimation of the muon zenith angle ($\theta_{w}$) and the corresponding parameter ($\theta_{t}$) of the reconstructed muon track by a neutrino telescope\footnote{This is the SeaWiet \cite{sw} neutrino telescope configuration, with an inter-string separation of 130m and an inter-OM separation of 40m, deployed in a depth of 3500m.}. 

This distribution corresponds to 39 hours of data collection with a single autonomous array. It consists of a central Gaussian structure (with a mean at $0.010^{o}$$\pm$$0.045^{o}$) but also exhibits and other spurious structures due to badly reconstructed tracks. These tracks can be eliminated by applying stringiest track selection criteria, at the expense of the available statistics.
It has been checked that this contamination of the event sample does not affect the resolution in investigating for angular offsets.

Similarly, we investigate for an offset in the azimuth angle reconstruction by comparing, event by event, the simple azimuth angle estimation with the corresponding angle of the reconstructed muon track. In Figure \ref{fig3} it is shown the distribution of the difference between the simple estimation of the muon azimuth angle ($\phi_{w}$) and the corresponding parameter ($\phi_{t}$) of the reconstructed muon track by the neutrino telescope. The statistical error in the position peak of the central Gaussian structure (corresponding to 39 hours of data taking) is 0.3$^{o}$ . 

Furthermore, one can estimate the absolute position of the detector by examining the distribution of the coordinates, with respect to the center of the autonomous array, of the point where the reconstructed track intercepts the horizontal plane, at sea level. However, the angular and spatial distributions described above are highly correlated. Our studies \cite{helycon} have shown that a spatial systematic displacement of the detector results in an apparent zenith angle distortion and vice versa, requiring thus careful analysis in order to disentangle the effects.

\section{Simulation and Results}
\label{DD}
In this paper we report on results for the SeaWiet \cite{sw} proposed telescope configuration consisting of 300 detection units (strings), each with 20 multi-PMT Optical Modules (OM), 130 m separation between strings and 40 m vertical separation between OMs\footnote{In a more extended note \cite{helycon} we present results for an extended SeaWiet layout as well as for the $\nu$One \cite{none} proposed configuration.}. We assumed that the telescope operates at a depth of 3500 m.

The CORSICA simulation package \cite{corsika} was used for the description of the EAS development, of primary particles (up to iron nucleus), of energy range  from 10 TeV up to 5 PeV. The HELYCON array response as well as the reconstruction of the shower parameters were performed by utilizing the HELYCON simulation and reconstruction software package \cite{idm,helycon}. For each EAS, producing energetic muon(s) which reach the underwater  detector, the response of the neutrino telescope was simulated and a full reconstruction analysis   was performed by utilizing the HOURS \cite{apohere} package. 

In this study a large statistical sample was simulated, corresponding to more than 8 million EAS. In order to study spatial effects the showers were spread randomly on an area of 500 m radius around an autonomous HELYCON array, positioned on top of the neutrino telescope’s center. The resolution of evaluating systematic offsets was defined as the statistical error in estimating the central Gaussian peaks, scaled to the expected number of events to be collected during a 10-day operation of three autonomous HELYCON arrays (henceforth full data set). These results have been checked by forming 2000 smaller sets of events (each corresponding to 39 hours operation of a single array, and containing, per average, 1950 events) by employing the statistical bootstrap technique \cite{helycon,boot}.  

\begin{figure}
\begin{center}
\includegraphics*[width=9cm]{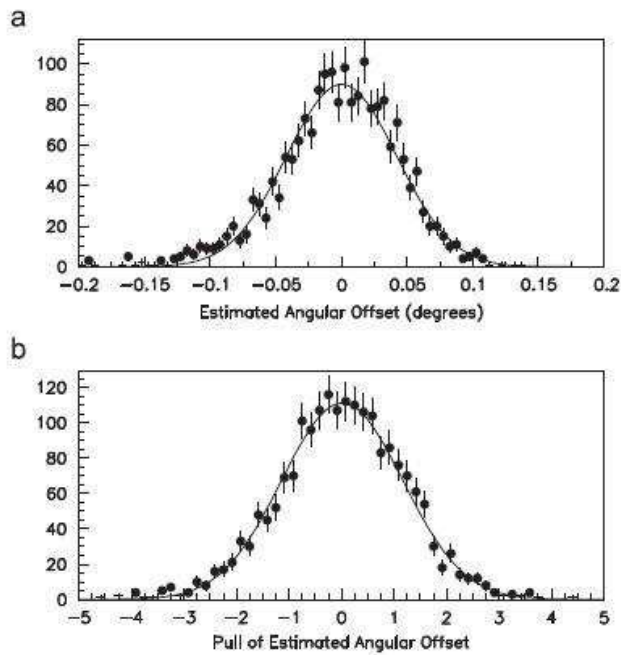}
\end{center}
\caption{ (a) Distribution of  the central peak positions, and (b) The Pull distribution of the central peak position estimation.}
\label{fig2}
\end{figure}

In Figure \ref{fig2} we present the estimations of the  central (Gaussian) peak positions, of the $\theta$$_{t}$-$\theta$$_{w}$ distributions for all the 2000 sets of simulated events. These estimations follow a Gaussian distribution, centered at 0, with a sigma of $0.042^{o}$ which by definition is the sensitivity in evaluating a possible angular offset. When the sensitivity is scaled to the full data set, the sigma falls to $1\cdot 10^{-2}$ degrees. Figure 2b demonstrates that the pull distribution of these peak positions (that is the estimated peak position divided by the estimation error) is indeed normal with a mean at zero and sigma consistent with one ($1.05\pm0.03$). This behaviour indicates that the statistical error in estimating the position of the central peak expresses indeed the resolution of the calibration technique.

Finally, our simulation studies conclude that the sensitivity in investigating for a systematic error in reconstructing the azimuth angle, corresponding to the full data set, is 0.07$^{o}$ and the sensitivity in estimating the x and y coordinates of the underwater detector is better than 1 m.

\section{Conclusions}
\label{CC}
We have studied a new strategy of using floating detector arrays in order to investigate for possible systematic errors in track reconstruction by an underwater neutrino telescope. Assuming that three floating arrays collect, independently of each other, data for a period of ten days, we found that a possible offset in zenith angle estimation can be evaluated with an accuracy of $0.01^{o}$, whilst a similar offset in the azimuth angle can be found with an acurracy of 0.07$^{o}$. Furthermore the coordinates of the center of the neutrino telescope, can be estimated with an accuracy better than  1 m. In this study we did not take into account correlations of the above systematic errors.

\section*{Acknowledgements}
This work is supported through the EU-funded FP6 KM3NeT Design Study Contract No. 011937.

\bibliographystyle{elsarticle-num}

\begin{thebibliography}{00}
\bibitem{Nestor}
G. Aggouras et al., Astroparticle Physics, Vol 23, 377-392, 2005.
\bibitem{apothesis}
A. Tsirigotis, PhD Thesis, Hellenic Open University 2004 , http://physicslab.eap.gr/thesis/tsirigotis , in Greek.
\bibitem{apocalib}
A. Tsirigotis et al.,  Nuclear Instruments and Methods in Physics Research A 595(2008) 80-23.
\bibitem{ernen}
J. P. Ernenwein et al, Nucl. Instrum. Meth. A 602 (2009) 88.
\bibitem{idm}
S.E. Tzamarias, “HELYCON: towards a sea-top infrastructure”, in: Proceedings of the 6th International Workshop on the Identification of Dark Matter (IDM 2006), World Scientific (2007), p. 464 (ISBN-13978-981-270-852-6).
\bibitem{bourlistot}
G.~Bourlis et al [KM3NeT Collaboration], Nucl.\ Instrum.\ Meth.\  A 602 (2009) 129 ; 10.1109/IWASI.2009.5184796 
\bibitem{bouthesis}
G. Bourlis, PhD Thesis, in prep., Hellenic Open University.
\bibitem{cdr}
KM3NeT ,Conceptual Design Report,,http://www.km3net.org
\bibitem{gaisser}
Gaisser T., “Performance of the IceTop Array”, Proceedings of the 30th International Cosmic Ray Conference.
\bibitem{apohere}
A. Tsirigotis et al, “HOU Reconstruction $\&$ Simulation (HOURS)” this issue.
\bibitem{sw}
SeaWiet write up, http://www.km3net.org
\bibitem{helycon}
``HELYCON", HOU Technical note to be published.
\bibitem{none}
$\nu$One write up, http://www.km3net.org
\bibitem{corsika}
J. Knapp, D. Heck, Nachr. Forsch. zentr. Karlsruhe 30 (1998) 27, http://www-ik.fzk.de/corsika
\bibitem{boot}
See for example, B. Efron, J. Am. Statist. Assc. 82(397).
\end{thebibliography}

\end{document}